# New Multielement Isotopic Compositions of Presolar SiC Grains: Implications for Their Stellar Origins

Nan Liu[1,2,3*], Jens Barosch[3], Larry R. Nittler[3], Conel M. O'D. Alexander[3], Jianhua Wang[3], Sergio Cristallo[4,5], Maurizio Busso[5,6], Sara Palmerini[5,6]

[1]Laboratory for Space Sciences and Physics Department, Washington University in St. Louis, St. Louis, MO 63130, USA; nliu@physics.wustl.edu

[2]McDonnell Center for the Space Sciences, Washington University in St. Louis, St. Louis, MO 63130, USA

[3]Earth and Planets Laboratory, Carnegie Institution for Science, Washington, DC 20015, USA

[4]INAF, Osservatorio Astronomico d'Abruzzo, Via Mentore Maggini snc, 64100 Teramo, Italy

[5]INFN, Sezione di Perugia, Via A. Pascoli snc, I-06123 Perugia, Italy

[6]Department of Physics and Geology, University of Perugia, Via A. Pascoli sns, I-06123 Perugia, Italy.

ABSTRACT

We report NanoSIMS Si and Mg-Al isotopic data (and C, N, and Ti isotopic data when available) for 85 submicron- to micron-sized presolar SiC grains from the CM2 Murchison meteorite, including 60 MS, 8 AB1, 8 X, 7 AB2, and 2 Y grains. The MS and Y grain data demonstrate that (1) C and N contamination mainly appears as surface contamination, and sufficient presputtering is needed to expose a clean grain surface for obtaining intrinsic C and N signals, and (2) Mg and Al contamination appears as adjacent grains and rims, and high-resolution imaging and the choice of small regions of interest during data reduction together are effective in suppressing the contamination. Our results strongly indicate that previous studies on presolar SiC grains could have sampled differing degrees of contamination for C, N, Mg, and Al. Compared to the literature data, our new MS and Y grains are in better agreement with carbon star observations for both the C and N isotopic ratios. By comparing our new, tighter distributions of $^{12}C/^{13}C$, $^{14}N/^{15}N$, and initial $^{26}Al/^{27}Al$ ratios for MS and Y grains with FRUITY AGB stellar models, we provide more stringent constraints on the occurrence of cool bottom processing and the production of $^{26}Al$ in N-type carbon stars, classical asymptotic giant branch stars.

1. INTRODUCTION

Presolar grains are microscopic solids that condensed around different types of stars at their advanced evolutionary stages before the solar system formation. As stellar condensates, presolar grains preserve a detailed record of nucleosynthesis and mixing processes in their parent stars and thus grant us the possibility of deriving a detailed picture of these stellar processes, if we can determine the types of stars they came from. Silicon carbide (SiC) is the best studied presolar mineral phase. There are multiple pieces of evidence that presolar SiC grains condensed dominantly from carbon stars. For instance, presolar SiC grains show a $^{12}C/^{13}C$ ratio distribution similar to that of carbon stars (Zinner 2014); SiC grains are predicted to condense in C-rich stellar environments according to thermodynamic equilibrium calculations (Lodders & Fegley 1995); SiC dust has also been observed to be ubiquitous in circumstellar envelopes around C-rich asymptotic giant branch (AGB) stars (e.g., Little-Marenin 1986; Chan & Kwok 1990; Yang et al. 2004). Presolar SiC grains were originally divided into five main groups (≥1% of the population), including mainstream (MS), X, Y, Z, and AB. MS grains were defined as SiC grains having $10 \lesssim {}^{12}C/^{13}C < 100$ and $\delta^{29}Si$[1] and $\delta^{30}Si$ correlating with a slope of ~1.35. In comparison, Y and Z grains have Si isotopic compositions deviating from the linear correlation observed for MS grains toward larger $^{30}Si$ enrichments relative to $^{29}Si$, with the deviations larger for Z grains than for Y grains. In addition, while MS and Z grains overlap in their C and N isotopic compositions, Y grains were defined to have $^{12}C/^{13}C \geq 100$. AB grains have low $^{12}C/^{13}C$ ratios ($\lesssim 10$) and a similar range of Si isotopic compositions as MS grains but exhibit a wider range of $^{14}N/^{15}N$ ratios. Finally, X grains exhibit large $^{15}N$ and $^{28}Si$ excesses and a wide range of $^{12}C/^{13}C$ ratios (see Zinner 2014 for a review). These classifications were based on C, N, and Si isotopic ratios mainly because, as the most abundant elements in the grains, they are the easiest to measure and these elements thus dominate the existing isotope database for presolar SiC. In fact, interpreting the C, N, and Si isotopic compositions of presolar SiC in the context of stellar nucleosynthesis is quite complex. Thus, linking the different groups to different types of carbon stars has proven to be quite a challenge (*e.g.*, Zinner et al. 2006, 2007). Boujibar et al. (2021; B21 hereafter) recently assessed the existing classification scheme by applying the technique of cluster analysis, and their results highlight the importance of $^{14}N/^{15}N$ and inferred initial $^{26}Al/^{27}Al$ ratios ($^{26}Al/^{27}Al$ hereafter, determined from $^{26}$Mg excesses) to presolar SiC grain classification and the need for more multielement isotopic data for more accurate classification.

---

[1] $\delta^{29}Si$ is defined as $[(^{29}Si/^{28}Si)_{grain}/(^{29}Si/^{28}Si)_{SS} - 1] \times 1000$, in which $/(^{29}Si/^{28}Si)_{SS}$ denotes the solar system value.

Multielement isotopic data for individual presolar SiC grains are limited. There are only ~100 MS grains with both $^{14}$N/$^{15}$N and $^{26}$Al/$^{27}$Al (< 50% errors) in the Presolar Grain Database (PGD; Hynes & Gyngard 2009; Stephan et al. 2021), although >10,000 MS SiC grains have been examined for their isotopic compositions. Moreover, asteroidal and terrestrial contamination complicates the data interpretation. Although the wide range of $^{14}$N/$^{15}$N ratios observed in MS grains may provide important constraints on astrophysical processes, it is possible that some or all of this large range is simply caused by differing amounts of N contamination sampled during laboratory analyses. Furthermore, evidence that Al contamination strongly affected $^{26}$Al/$^{27}$Al data in the literature was presented by Groopman et al. (2015). For the case of some trace-element measurements, *e.g.*, Sr and Ba, special cleaning procedures have been developed (e.g., Liu et al. 2015), but similar methods have not been applied to studies of important elements like N, Mg, and Al.

In this study, we examined the problem of contamination by adopting two approaches. For C, N, and Si isotopes, we conducted NanoSIMS analyses on the same grains after (*i*) minimal presputtering (a few second to minute sputtering for the Si$^-$ count rate to reach equilibrium) and (*ii*) extensive presputtering for comparison (see Appendix A for details), which enabled us to determine the source and degree of contamination for these elements. For Mg-Al, Ca-Ti, and Ti-V isotopic systematics, we used an RF plasma primary ion source, the Hyperion (Oregon Physics, LLC), which produces an O$^-$ beam of reduced size compared to previous duoplasmatron ion sources (Malherbe et al. 2016). The improved spatial resolution enabled us to suppress contamination from the surrounding substrate for these elements.

## 2. RESULTS

All the isotopic data were collected with the Cameca NanoSIMS 50L instrument at the Carnegie Institution in imaging mode; the analytical details are given in Appendix A. We obtained Si and Mg-Al isotopic data for all 85 grains investigated (Table 1) and Ti isotopic data for 35 of the grains (Table A1). Prior to these isotopic analyses using the Hyperion O$^-$ source, we used a Cs$^+$ ion source to obtain C, N, and Si isotopic ratios (hereafter pre-Hyperion data) for all AB and X grains[2] and 25 of the MS and Y grains. For 47 of the MS and Y grains, we also collected their C and N isotopic data after the Hyperion analyses in the remaining grains (hereafter post-Hyperion data). The C and N isotopic ratios and $^{26}$Al/$^{27}$Al are reported as absolute ratios, while Mg, Si, and Ti isotopic ratios are reported in δ notation, defined as $\delta^i A = ((^iA/^jA)_{grain}/(^iA/^jA)_{standard} - 1) \times 1000$, in which $(^iA/^jA)_{grain}$ and $(^iA/^jA)_{standard}$ denote the measured ratios of a grain and standard, respectively; the normalizing isotopes for Mg, Si, and Ti are $^{24}$Mg, $^{28}$Si, and $^{48}$Ti, respectively. All data are reported with 1 σ errors.

A comparison of the pre-Hyperion, post-Hyperion, and PGD data (PGD_SiC_2021-01-10; Stephan et al. 2021) for MS, Y, and Z grains is given in Fig. 1. We chose these three types of grains for comparison because MS grains are the dominant type (Zinner 2014) and the three types are all inferred to have come from low-mass AGB stars given their comparable *s*-process Mo isotopic signatures (Liu et al. 2019). Figure 1a shows that (1) the PGD data have a distribution (in grayscale colormap) that peaks at $^{12}$C/$^{13}$C ≈ 50–55 and $^{14}$N/$^{15}$N ≈ 500, (2) our pre-Hyperion data for 25 MS and Y grains (Table A1) overlap with the PGD data (our average $^{12}$C/$^{13}$C = 59.5, $^{14}$N/$^{15}$N = 756), and (3) the post-Hyperion data for the same grains (Table 1) show significantly enhanced $^{14}$N/$^{15}$N ratios with the average being shifted upward to ~2000. The significant increase in $^{14}$N/$^{15}$N between the pre- and post-Hyperion data suggests that N contamination was present across the whole grain surface during the pre-Hyperion measurements, which could not be effectively suppressed even by selecting small regions of interest (ROIs) with low N within the grain[3] during the data reduction; instead, extensive sputtering was needed to expose clean grain surfaces for obtaining intrinsic N signals. Differences in $^{12}$C/$^{13}$C between the pre- and post-Hyperion data are overall small – the averages are 59.5 and 56.7, respectively. The probability distribution of all our 47 MS and Y grain data (post-Hyperion data in Table 1) is shown in Fig. 1b for comparison with the PGD data in Fig. 1a. Given its absence in our post-Hyperion data, the secondary peak at $^{12}$C/$^{13}$C ≈ 90 and terrestrial/solar-like $^{14}$N/$^{15}$N ratios for the PGD data, is most likely the result of significant asteroidal/terrestrial contamination.

---

[2] We did not determine the $^{14}$N/$^{15}$N ratio for three of the X grains due to a problem with one of the detectors during that analytical session.
[3] The grain contour is inferred from the $^{12}$C and/or $^{28}$Si ion images (see Fig. A1 for example).

Our MS and Y grains show $^{26}$Mg/$^{24}$Mg values that are more than one order of magnitude higher than those of Huss et al. (1997)[4]. Our higher $^{26}$Mg/$^{24}$Mg values result from sampling less Mg contamination, as evidenced by our lower Mg/Si (Fig. 2b) and enhanced Al/Mg ratios. The Huss et al. data and our data lie close to the same trendline in Fig. 2a. Hence, the differences in the Al/Mg ratios between the two studies cannot be explained by higher levels of Al contamination in our analyses because this would have moved our data horizontally to the right of the trendline in Fig. 2a. Thus, Figs. 2a,b together point to that our study sampled much reduced Mg contamination, which was made possible by both the high spatial resolution achieved with the Hyperion O$^-$ beam, and the use of ROIs that were smaller than the full grains to exclude adjacent Mg contamination (see Fig. A1 and Appendix A for details). This was not possible with the ion probe technology of the 1990s.

The high spatial resolution and exclusion of Al contamination (Fig. A1) also resulted in higher $^{26}$Al/$^{27}$Al ratios, calculated by assuming that all $^{26}$Mg excesses were produced by the decay of $^{26}$Al (see Appendix A for details), for our MS and Y grains (Fig. 2c). We simulated the effect of having a larger primary beam on the levels of Al contamination by recalculating the $^{26}$Al/$^{27}$Al ratios while including all Al signals within each grain (blue circles in Fig. A2). This lowered $^{26}$Al/$^{27}$Al ratios by up to 60% (45% on average). The effects of Mg and Al contamination on reducing the Mg isotopic ratios and $^{26}$Al/$^{27}$Al ratios, respectively, of grains from Huss et al. (1997) are further illustrated in Fig. A3.

Our high-resolution ion images also revealed (1) common Ca and Cr contamination present on the substrate, and (2) insignificant Si, Ti, and V contamination (Fig. A1). The literature MS grains have an average $\delta^{50}$Ti of 179±75‰ (one standard deviation) and our MS grains 153±95‰. In agreement with the literature data, our AB grains exhibit less correlated Si and Ti isotopic compositions than the MS and Y grains. For instance, $\delta^{46}$Ti and $\delta^{29}$Si show a strong correlation among MS and Y grains (with Pearson's correlation coefficient $R = 0.79$ and 0.78 for the PGD and our data, respectively), but a weaker correlation among AB grains (with $R = 0.58$ and 0.44, respectively).

---

[4] We chose to compare our data with only those of Huss et al. (1997) because (1) the sizes of the two sets of data are similar, and (2) many other studies reported only the inferred $^{26}$Al/$^{27}$Al ratios but not the measured Mg isotopic data.

In summary, we identified substantial N and Al contamination, which has certainly impacted the literature data by, for example, lowering $^{14}$N/$^{15}$N and initial $^{26}$Al/$^{27}$Al ratios in the PGD and thus compromising inferred ratio distributions. In the next section, our discussion focuses on the grain classification and AGB stellar nucleosynthesis based on our new $^{14}$N/$^{15}$N and $^{26}$Al/$^{27}$Al results for MS, Y, and AB grains.

## 3. DISCUSSION AND CONCLUSIONS

### 3.1 Comparison with Observations for Carbon Stars

Stellar nucleosynthesis model calculations suffer from uncertainties in the initial stellar composition, non-standard mixing processes, and nuclear reaction rates (see Busso et al. 1999; Palmerini et al. 2011 for reviews). Given the challenges in using model calculations to explain presolar grains' C and N isotopic ratios, we first compare the MS, Y, and AB grain data with observations for carbon stars (Table 2) in Fig. 3a. N-type carbon stars represent classical AGB stars and are the most numerous type of carbon star in the Galaxy. Most N-type carbon stars with near-solar initial metallicities are enhanced in *s*-process elements and F, which results from the recurrent mixing of He-burning products from the stellar interior into the envelope through third dredge-up (TDU) episodes (Abia et al. 2001, 2008; Abia 2008). J-type carbon stars represent about 10–15% of all Galactic carbon stars (Abia & Isern 2000; Morgan et al. 2003) and are characterized by peculiar chemical compositions, including low $^{12}$C/$^{13}$C ratios, Li enhancements (80% of the stars), and no *s*-process element (*e.g.*, Sr, Zr, Ba, Nd) or F enhancements (Abia & Isern 2000; Abia et al. 2015). The origin of J-type carbon stars is unknown. SC-type carbon stars are rare in the Galaxy (Abia & Wallerstein 1998). They are characterized mainly by C/O ratios of around unity and are chemically similar to N-type carbon stars. In the classical picture, a SC-type carbon star represents a brief AGB phase when an oxygen-rich, class-M star transitions into a carbon star as TDU episodes bring increasing amounts of $^{12}$C to the stellar surface. N- and SC-type carbon stars in the solar neighborhood are inferred to be 1.5–3 $M_\odot$ AGB stars according to their derived luminosity functions (Abia et al. 2020).

### 3.1.1 MS/Y grains and N-, SC-type carbon stars

The grain-observation comparison in Fig. 3a links MS and Y grains to N-type carbon stars, classical low-mass AGB stars. MS and Y grains make up more than 90% of presolar SiC grains, which is consistent with the observation that N-type carbon stars are the large majority of carbon stars in the Galaxy and in the Large Magellanic Cloud (e.g., Lloyd Evans 2010). Our MS and Y grain C and N isotopic data are very similar to those of N-type carbon stars in Fig. 3a, *i.e.*, the stars mostly lie within the region where most of the grain data plot (Fig. A4). The consistency implies that N contamination was effectively suppressed in our analyses, such that the measured N (and C) isotopic ratios are representative of the envelope compositions of the grains' parent stars. Thus, our results solve a longstanding discrepancy highlighted by many previous studies (e.g., Hedrosa et al. 2013), in which $^{14}$N-poor MS grains could not be reproduced by theoretical AGB stellar models and were linked to astrophysical sites other than N-type carbon stars.

Despite the general consistency, a slight difference is seen in the range of $^{14}$N/$^{15}$N: while N-type stars generally have $^{14}$N/$^{15}$N above 1000, some of our MS grains show $^{14}$N/$^{15}$N down to the terrestrial value. This raises the question of whether these grains represent SiC from a different type of star, or their lowered ratios were caused by sampling N contamination. B21 pointed out that MS grains with $^{14}$N/$^{15}$N $\lesssim$1000 ($^{15}$N-rich MS grains hereafter; brown ellipse in Fig. 3b) tend to show higher $^{26}$Al/$^{27}$Al than those with high $^{14}$N/$^{15}$N ($^{14}$N-rich MS grains hereafter; blue ellipse), on the basis of which they divided MS grains into two groups. In agreement with the B21 results, $^{26}$Al/$^{27}$Al $\gtrsim 3\times10^{-3}$ was found in four of our seven $^{15}$N-rich MS grains (57.1%), but in none of our 33 $^{14}$N-rich MS/Y grains. However, Fig. 3b shows that the definition for these two clusters given by B21 based on the PGD is not applicable to our new grain data due to the PGD's N and Al contamination problem.

Figure 3a suggests that the $^{15}$N-rich MS grains may have come from SC-type stars. The rareness of SC-type stars (3% of all AGB stars) naturally explains the scarcity of $^{15}$N-rich MS grains. Abia et al. (2017) observed that, compared to N-type carbon stars, SC-type carbon stars tend to exhibit higher $^{17}$O/$^{16}$O ratios, leading to the suggestion that SC-type carbon stars could have higher initial stellar masses ($\gtrsim$3–4 $M_\odot$) than typical N-type carbon stars ($\lesssim$3 $M_\odot$). However, the stellar models of Karakas (2014) for 3–4 $M_\odot$ carbon stars (compared to $\lesssim$3 $M_\odot$ carbon stars) predict enhanced $^{14}$N production and no $^{26}$Al/$^{27}$Al increase at the stellar surface (Karakas & Lugaro 2016), and thus cannot explain the $^{15}$N- and $^{26}$Al-rich nature of $^{15}$N-rich MS grains. Instead, these minor isotope enrichments, together with the observed high $^{17}$O/$^{16}$O ratios for SC-type carbon stars, point to the occurrence of high-temperature ($\gtrsim 1\times10^8$ *K*) H-burning process(es), *i.e.*, nova-like nucleosynthesis (José et al. 2004), in SC-type stars, which are not predicted by any of the state-of-the-art stellar models for 1.5–4 $M_\odot$ carbon stars and need further investigation.

*3.1.2 AB grains and J-, SC-type carbon stars*

Type AB grains with higher-than-solar $^{14}N/^{15}N$ ratios are consistent with observations of both J-type carbon stars and a few SC-type carbon stars in Fig. 3a. AB grains were defined as having $^{12}C/^{13}C \lessgtr 10$ and a wide range of $^{14}N/^{15}N$ ratios. Based on differences in $^{26}Al/^{27}Al$, Ti, and Mo isotopic ratios, Liu et al. (2017a, 2017b, 2018a) proposed to further divide AB grains into two subgroups, AB1 (with subsolar $^{14}N/^{15}N$) and AB2 (with higher $^{14}N/^{15}N$), using the solar $^{14}N/^{15}N$ as a divider. This division scheme is supported by the observed $^{14}N/^{15}N$ ratios of SC-type and J-type carbon stars. The B21 cluster analysis also suggested that there are two clusters of AB grains but with some overlap in their $^{14}N/^{15}N$ ratios (ellipses in Fig. 3a). Liu et al. (2017b) proposed that AB2 grains mostly came from J-type carbon stars based on their general lack of *s*-process Mo isotopic signatures. However, two of the 12 AB2 grains from Liu et al. (2017b) and one AB2 grain from Stephan et al. (2019) exhibit *s*-process Mo isotopic signatures and are consistent with an origin in SC-type stars.

The proposal that AB1 grains originated in supernovae that experienced explosive H burning in the He/C zone (Liu et al. 2017a, 2018a), is supported by the stellar observations in Fig. 3a given the lack of carbon stars with $^{14}N/^{15}N<440$ and $^{12}C/^{13}C\leq10$. Schmidt et al. (2018) reported large $^{13}C$, $^{17}O$, and $^{15}N$ enrichments in a young, carbon-rich planetary nebula (K4-47), based on which the authors proposed that the precursor of the nebula was a J-type star that underwent a He-shell flash. The result of Schmidt et al. (2018) raises the possibility some J-type carbon stars have subsolar $^{14}N/^{15}N$ ratios. However, K4-47 could represent an extended nebula that was ejected by a pair of interacting binary stars during a nova-like explosion (Corradi et al. 2000), which is supported by the observation of high-velocity bullets of material ploughing through the surrounding interstellar medium (e.g., Gonçalves et al. 2004). Moreover, such a low $^{14}N/^{15}N$ ratio as observed by Schmidt et al. (2018), 13.6±6.5, has yet to be observed for any J-type carbon stars (Hedrosa et al. 2013). Thus, $^{15}N$-rich J-type carbon stars are relatively rare if they exist at all. In contrast, AB1 and AB2 grains are about equally abundant and, unlike AB2 grains, AB1 grains generally exhibit moderate *s*-process isotopic enrichments (Liu et al. 2018a). Since J-type carbon stars are characterized by the lack of any *s*-process enrichments, J-type carbon stars are unlikely to be the sources of AB1 grains.

### 3.2 Comparison with FRUITY Stellar Models for Low-mass AGB Stars

As discussed above, the $^{14}$N-rich MS and Y grains likely represent dust from N-type carbon stars, *i.e.*, 1.5–3 $M_\odot$ AGB stars with close-to-solar metallicities (Abia et al. 2020). Here we compare the isotopic data for these grains as well as the measurements of N-type carbon stars to appropriate FRUITY[5] stellar models in Figs. 3c, 3d. The FRUITY stellar models were computed by fully coupling a full nuclear network to the updated FUNS stellar evolution code[6] (see Cristallo et al. 2009, 2011 for details), in contrast to the postprocessing approach commonly adopted in the literature. Besides the use of a full nuclear network, FUNS differs from previous versions of the code in other important details. In particular, FUNS adopts low-temperature molecular opacities that take into account the increased opacity due to the formation of C-bearing and N-bearing molecules at low temperatures (T < 5000 K) (Cristallo et al. 2007). Moreover, the adopted mass-loss rate is calibrated against the physical properties of a sample of Galactic giant stars (see Straniero et al. 2006 for details). The FRUITY stellar models predict higher $^{12}$C/$^{13}$C and $^{26}$Al/$^{27}$Al ratios and a narrower range of $^{14}$N/$^{15}$N ratios than are present in the grain data. Galactic chemical evolution (GCE) could have caused the initial compositions of the grains' parent stars to deviate from the assumed solar isotopic starting compositions (except for $^{14}$N/$^{15}$N [7]) adopted in the FRUITY model calculations. However, this is unlikely because the parent stars of the MS and Y grains were born at much earlier times than the observed N-type carbon stars, yet the MS and Y grains and the N-type carbon stars show comparable C and N isotopic ratios. Instead, the inconsistent C and N isotopic ratios between observations and models are probably the result of cool bottom processing (CBP) in N-type carbon stars. CBP is not taken into account in the FRUITY models (shown as lines with symbols in Fig. 3) and can enhance a star's $^{13}$C, $^{14}$N, and $^{26}$Al production as envelope material is circulated to deep regions where H-burning reactions can take place at enhanced stellar temperatures (Wasserburg et al. 1995). The cause of this circulation, which may already work during the red giant branch (AGB) phase, is as yet unknown, but may be related to magnetic buoyancy effects (Palmerini et al. 2017).

---

[5] FRUITY stands for FRANEC Repository of Updated Isotopic Tables & Yields (Cristallo et al. 2011, 2015). The FRUITY database is available at http://fruity.oa-teramo.inaf.it/.

[6] FUNS code (Straniero et al. 2006) is a more recent version of the original FRANEC code (Chieffi & Straniero 1989). As already highlighted, the main novelty of the FUNS code is the adoption of a full nuclear network, which involves about 500 isotopes linked by more than 1100 reactions that are directly coupled to the physical evolution of the stellar structure (see e.g., Cristallo et al. 2009). For a full description of the FUNS code, we refer the reader to Straniero et al. (2006).

[7] For the N isotopic composition, the FRUITY models adopted the terrestrial value, 272, in contrast to the solar value, 440 (Marty et al. 2011). Thus, adopting the solar value would enhance the $^{14}$N/$^{15}$N ratios predicted by the FRUITY models. Note that the initial composition of the grains' parent stars (older than the Solar System) is largely unknown due to uncertainties in the GCE.

Given the uncertainties in the underlying mechanism for CBP, Palmerini et al. (2011) implemented CBP in the 2 $M_\odot$, $Z_\odot$ FRUITY stellar model using two parameters (shown as red lines in Fig. 3c): (1) maximum circulation temperature (the first number next to the red lines), which is defined as $\Delta = \log T_H - \log T_P$ where $T_H$ is the T at which the energy from the H-burning shell is maximum ($6.3 \times 10^7$ K) and $T_P$ the maximum T sampled by circulating material, and (2) transport rate (the second number, in the unit of $10^{-6}$ $M_\odot$/year). A comparison of the grain data with the CBP calculations in Fig. 3c constrains the upper limits for the rate of mass transport and the maximum temperature sampled by the circulating material to $3.5 \times 10^7$ K and $3 \times 10^{-7}$ $M_\odot$/year, respectively. Within this parameter regime, the CBP calculations predict negligible changes in the $^{26}$Al production. Thus, after the CBP is taken into consideration, two problems remain: (1) none of the models can reach $^{14}$N/$^{15}$N≲1000, but a dozen of our MS grains lie within this region (Fig. 3c), and (2) the models generally predict higher $^{26}$Al/$^{27}$Al ratios than the grain data (Fig. 3d).

First, the fact that the minimum $^{14}$N/$^{15}$N predicted by FRUITY models for low-mass carbon-rich AGB stars lies above ~1000, supports our earlier conclusion that $^{15}$N-rich MS grains did not come from classical low-mass AGB stars (section 3.1.1). Second, the lower-than-predicted $^{26}$Al/$^{27}$Al ratios inferred for most of our MS and Y grains are unlikely to have been caused by Al contamination. Figure A5 indeed suggests that Al contamination was not completely suppressed for grains with diameters ≲500 nm (in green shaded area), resulting in their lowered $^{26}$Al/$^{27}$Al ratios. However, among larger SiC grains no dependence of $^{26}$Al/$^{27}$Al on grain size can be seen (Fig. A5), and we observed homogeneous $^{26}$Al/$^{27}$Al ratios in the cores of these grains (see Fig. A1 for example), pointing to negligible Al contamination present in our selected ROIs. Thus, our MS and Y grain data reveal that the $^{26}$Al/$^{27}$Al ratios at the surface of their parent AGB stars mainly fall within the range of $(1-2) \times 10^{-3}$ (Fig. 2c), in contrast to the FRUITY model predictions that lie above $2 \times 10^{-3}$ during the C-rich phase (Fig. 3d).

The Astrophysical Journal LettersThe data-model discrepancies in $^{26}$Al/$^{27}$Al ratios are most likely caused by uncertainties in the $^{26}$Al$^g$($p,\gamma$)$^{27}$Si reaction rate. To test this, we adopted the upper limit for this reaction rate given by Iliadis et al. (2010) in our 2 $M_\odot$, $Z_\odot$ stellar model, which reduces the $^{26}$Al/$^{27}$Al ratio by a factor of two during the C-rich phase (*e.g.*, from 4.27×10$^{-3}$ to 2.11×10$^{-3}$ at the last thermal pulse) and thus leads to a better grain-model agreement. On the other hand, (1) adoption of the new $^{25}$Mg($p,\gamma$)$^{26}$Al reaction rate estimated based on LUNA laboratory measurements (Straniero et al. 2013) resulted in only a ~15% decrease in the predicted $^{26}$Al/$^{27}$Al ratio, and (2) although possible enhancements of the $^{26}$Al($n,p$)$^{26}$Mg and $^{26}$Al($n,\alpha$)$^{23}$Na reaction rates could lower the $^{26}$Al/$^{27}$Al ratio, data obtained by the n_TOF experiment (Guerrero et al. 2013) point to rate changes that are in the wrong direction (Lederer et al. 2021).

In conclusion, a comparison of our MS, Y, and AB grain data with observations for carbon stars suggests that $^{14}$N-rich MS and Y grains originated in N-type carbon stars, $^{15}$N-rich MS and Y grains in $^{12}$C-rich SC-type carbon stars, and AB2 grains in J-type and $^{13}$C-rich SC-type carbon stars. The lack of observed carbon stars with $^{14}$N/$^{15}$N<440 and $^{12}$C/$^{13}$C≤10 supports that AB1 grains with such isotopic signatures condensed from ejecta of explosive events. Compared to $^{14}$N-rich MS and Y grains from N-type stars, $^{15}$N-rich MS and Y grains are more enriched in $^{26}$Al, and AB2 grains are more enriched in $^{13}$C and $^{26}$Al, both of which point to the occurrence of high-temperature H-burning process(es) in their parent stars. Finally, we highlight the urgent need for more multielement isotopic data for AB2 and $^{15}$N-rich MS, Y, and Z grains, especially their coupled $^{14}$N/$^{15}$N, initial $^{26}$Al/$^{27}$Al, and heavy-element isotopic compositions, by adopting the analytical procedures for suppressing contamination for these elements. This will shed more light onto the mysterious evolution of SC-type and J-type carbon stars.

Acknowledgements: We would like to thank Dr. Diego Vescovi for his help with running some of the FRUITY model tests for this study. We also would like to thank Dr. Ryan Ogliore for pointing out a small mistake in the manuscript. This work was supported by NASA through grants 80NSSC20K0387 to N.L. and NNX10AI63G and NNX17AE28G to L.R.N.**Appendix A**

Analytical Methods

**Samples**

The SiC grains in this study were extracted from the CM2 carbonaceous chondrite Murchison by using the isolation method described in Nittler & Alexander (2003). The extracted SiC grains have a median diameter of 0.9 μm. Isotopic data for a number of AB, Y, Z, X, putative nova, and ungrouped grains on the prepared mounts have been previously reported (Liu et al. 2017a,b, 2018a,b, 2019). The presolar SiC grains investigated in this study were identified by both scanning electron microscope (SEM)-based energy dispersive X-ray (EDX) analyses and isotopic analyses of C, N, and Si with the Cameca NanoSIMS 50L instrument at the Carnegie Institution. A total of 85 new SiC grains were chosen for NanoSIMS isotopic analyses in this study, including 60 MS, 15 AB, 8 X, and 2 Y grains. Note that grains on mount #4 (named M4-AX-X) were first measured for Mg-Al and Si isotopes, during which 17 of the 41 SiC grains were completely consumed and could not be further measured for their C and N isotopic compositions. While the Si isotopic data clearly indicate that two of the 17 grains are X grains given their large $^{28}$Si depletions, the Si isotopic data alone cannot tell MS and AB grains apart. Also, the analytical uncertainties in the measured Si isotopic ratios are relatively large for the other 15 grains, thus precluding us to distinguish between MS, Y, and Z grains. However, given the dominance of MS grains among presolar SiC grains (>85%), the inferred stellar sources for these grains are likely robust, considering that MS, Y, and Z grains all came from low-mass AGB stars and that less than one grain (by assuming 5% AB grains among presolar SiC) is expected to be of type AB. Thus, the grain-type assignment for these 15 grains solely based on Si isotopic ratios is not expected to affect any of the discussions in the main text. In addition, we chose 25 well-isolated (*i.e.*, no adjacent grains), relatively large (0.84 μm on average) MS and Y grains on mounts #1, 2, 3 to investigate the problem of C and N contamination by collecting pre-Hyperion and post-Hyperion C and N isotopic data (see next section for details).

Note that Figure 3a suggests that the MS grain M1-A4-G508 ($^{12}$C/$^{13}$C=13.3±0.3) could be classified as an AB2 grain, as this grain overlaps well with one of the J-type stars. The pre-Hyperion analysis for this grain yielded a $^{12}$C/$^{13}$C ratio of 19.4±0.1 (Table A1). Also, note that the AB grain data in Fig. 3a are all pre-Hyperion data, and the true $^{14}$N/$^{15}$N ratios of the AB2 grains are probably a factor of three higher on average (see Section 2 for discussion), which explains that the AB2 grains generally show lower $^{14}$N/$^{15}$N ratios than observed in J-type carbon stars (Fig. A4) and consequently implies that some of the AB1 grains with terrestrial≤$^{14}$N/$^{15}$N<solar could be AB2 grains instead.

**NanoSIMS Analytical Procedure**

All isotopic analyses were conducted with the Cameca NanoSIMS 50L instrument at the Carnegie Institution. For 48 grains, a Cs$^+$ primary ion beam of ~1−3 pA (16 keV energy) was first used for isotopic analyses of C, N, and Si as negative ions (N measured as CN$^-$) at a spatial resolution of ~100−150 nm (pre-Hyperion data for AB and X grains in Table 1 and those for MS/Y grains in Table A1). After that, a primary O$^-$ ion beam of ~3−10 pA produced by the Hyperion source was used here for Mg/Al, Ca, and Ti isotopic analyses, resulting in high spatial resolution (~150-300 nm) in the obtained ion images. Similar to our previous analytical procedure (Liu et al. 2018b), we conducted Ti isotopic analysis in the multicollection mode instead of the combined analysis mode to improve the analytical stability and reliability. We measured Ca-Ti and Ti-V isotopic ratios in two sets of runs in the multicollection mode: $^{28}$Si$^+$, $^{40}$Ca$^+$, $^{48}$Ti$^+$, $^{49}$Ti$^+$, $^{50}$Ti$^+$, $^{51}$V$^+$, $^{52}$Cr$^+$ in the first set and $^{28}$Si$^+$, $^{40}$Ca$^+$, $^{44}$Ca$^+$, $^{46}$Ti$^+$, $^{47}$Ti$^+$, $^{48}$Ti$^+$, $^{51}$V$^+$ in the second set of runs. For N, the high $^{14}$N/$^{15}$N ratios obtained with IMS-3f (Hoppe et al. 1996; Huss et al. 1997) may suggest that extensive pre-sputtering is an effective way to suppress N contamination. This is because, with respect to corresponding NanoSIMS measurements, these old IMS-3f measurements had to consume more material, as the IMS-3f instrument is equipped with only one detector (five or seven detectors on NanoSIMS) and has a lower transmission efficiency. Thus, after the Hyperion-NanoSIMS analyses, we switched back to the Cs$^+$ source and measured the C, N, and Si isotopic ratios in all the remaining grains (post-Hyperion data for MS/Y grains in Table 1). The hope was that the additional sputtering of the sample and surrounding substrate would have reduced any contamination if present. Note that during the post-Hyperion analyses, there was a problem with one of the detectors, and we chose to collect $^{12}$C$^{12}$C$^-$, $^{12}$C$^{13}$C$^-$, $^{12}$C$^{14}$N$^-$, $^{12}$C$^{15}$N$^-$, $^{28}$Si$^-$, and $^{29}$Si$^-$, which means that the $^{30}$Si/$^{28}$Si ratio could not be determined in the post-Hyperion analyses.

All the isotopic data were collected in imaging mode, which allowed for the selection of smaller ROIs for calculating isotopic ratios and thus reduced the likelihood for contamination (see Fig. A1 for an example). The measurements were all done with an entrance slit width of 30 μm, exit slit widths of 50 μm (minor isotopes, e.g., $^{13}$C) and 80 μm (major isotopes, e.g., $^{12}$C), and all the secondary ions were counted with electron multipliers. The achieved mass resolving powers (5000−8000) were sufficient to suppress all potential molecular isobaric interferences, e.g., resolving $^{12}$CH from $^{13}$C at mass 13. Detector background was found to be between 0.001 and

The Astrophysical Journal Letters

0.01 cps, which compares with the lowest ion signal intensity of 0.3 cps for $^{25}$Mg$^+$ in grain M1-A5-G960.

To demonstrate the effectiveness of our two approaches (extensive presputtering and reduced ROIs) in suppressing potential contamination, the two sets of $^{14}$N/$^{15}$N (pre-Hyperion versus post-Hyperion) and $^{26}$Al/$^{27}$Al (all Al signals included versus Al contamination excluded) ratios are compared in Fig. A2 for MS/Y grains from this study. The Mg-Al isotopic systematics of our MS and Y grains are also compared with those from Huss et al. (1997) for their distribution in Fig. A3, from which we infer significant Mg and Al contamination sampled in the study of Huss et al. (1997). Although presolar SiC grains are generally free of Ti contamination (Figs. A1, A4), the presence of Cr can result in an isobaric interference at mass 50, thus compromising the analysis of $\delta^{50}$Ti. Since Ti and Cr signals do not always overlap in ion images of presolar SiC grains (Fig. A1), the improved spatial resolution here allowed us to suppress the Cr contribution by choosing small ROIs to exclude Cr contamination and thus to derive more reliable $\delta^{50}$Ti values ($^{50}$Cr contribution generally <10% with an average of 5%). The high-spatial resolution achieved by the Hyperion source also enabled direct observation of Ti,V-rich subgrains within the SiC host in NanoSIMS ion images (Fig. A6), which had mainly been previously achieved by transmission electron microscopic (TEM) observations. The 1σ errors reported in all tables and figures include both analytical (estimated based on 1σ standard deviations of all standard measurements) and counting statistical uncertainties (1σ Poisson errors).

A variety of standards with terrestrial isotopic compositions were used as isotopic standards and also for determining the relative sensitivity factor (RSF), defined as $(A/B)_{true}$ = $(A/B)_{measured}$/RSF, in which A and B represent two elements of interest. Synthetic SiC and Si$_3$N$_4$ grains were used as standards for C, N, and Si isotopic analyses. The NIST glass standard, SRM 610, was used as a standard for Ca-Ti and Ti-V isotopic analyses, while Burma spinel was used as a standard for Mg-Al isotopic analysis. For Ti isotopic analysis, synthetic TiC was used as an additional standard. The sensitivity factors of Mg, Ca, and Ti relative to Si were determined by measuring SRM 610, and the obtained values were 3.10, 9.78, and 4.02, respectively. The RSF of Mg/Al was determined to be 1.20 in Burma spinel for deducing $^{26}$Al/$^{27}$Al. The $^{26}$Al/$^{27}$Al ratios were determined from the collected ion counts by use of the equation $^{26}$Al/$^{27}$Al = [$^{26}$Mg$_{measured}$ − $^{24}$Mg$_{measured}$ × ($^{26}$Mg/$^{24}$Mg)$_{standard}$]/($^{27}$Al$_{measured}$ × RSF), where ($^{26}$Mg/$^{24}$Mg)$_{standard}$ is the measured

$^{26}$Mg/$^{24}$Mg ratio of the Burma spinel standard (*e.g.*, Nittler et al. 2008). However, it is questionable if O-rich standards such as SRM 610 are appropriate for determining RSF values for C-rich samples like SiC, and we examine this problem in Appendix B by comparing the NanoSIMS elemental data of presolar SiC grains with the corresponding EDX data obtained with an SEM and TEM when available.

## Appendix B

Mg/Si and Al/Si RSF Values for SiC

Although the SIMS ionization efficiency for positive ions shows a strong correlation with the ionization potential, variations are expected depending on the sample matrix and on the element itself. For example, the presence of oxygen in the sample enhances positive ion yields for most elements, and an O$_2$ flooding gun was, therefore, used in old 3f-IMS analyses to boost the yields of Mg$^+$, Al$^+$, and Ti$^+$ ions for presolar SiC measurements (e.g., Hoppe et al. 1996; Huss et al. 1997). However, the NanoSIMS is not equipped with such an O$_2$ flooding gun. The contrasting matrix chemistries between the standard SRM 610 (silicon glass) and SiC raises the question of whether RSF values determined based on SRM 610 measurements can be directly applied to calculate trace element abundances (Mg, Al, K, Ca, Ti, V, Cr) for SiC samples. In comparison to NanoSIMS analyses, SEM-EDX trace-element analyses are less affected by differing sample-standard matrix chemistries but more affected by other factors such as topography. We, therefore, compared the RSF-corrected NanoSIMS Mg and Al data for SiC grains from this study and Liu et al. (2018b) with their SEM-EDX data when available. The EDX data were obtained with a JEOL 6500F field emission SEM equipped with an Oxford Instruments silicon drift detector in auto-particle analysis mode (Liu et al. 2017c). The data were quantified by using pyrope (Mg,Al garnet) as the standard for Si and ENAL20 (enstatite aluminum 20 wt%, an Al-bearing enstatite glass) for Mg and Al. In addition, EDX data were also obtained with a TEM on electron transparent thin sections that had been extracted from X grains M2-A1-G674 and M2-A2-G1036 by using a focused ion beam instrument (Singerling et al. 2021). It is noteworthy that the Mg/Si ratios in MS SiC grains are determined by NanoSIMS to be <0.003 (e.g., Hoppe et al. 2016; Table 1), and Mg in single presolar SiC grains is, therefore, below the detection limit of SEM-EDX analyses. The high Mg abundances reported for X grains in Fig. A7b are dominated by $^{26}$Mg from $^{26}$Al decay, since most X grains are characterized by extremely high $^{26}$Al/$^{27}$Al ratios ($\gtrsim$0.1; see Liu et al. 2017c for details). Errors in the trace element data are not reported in Fig. A7, because the associated statistical errors are negligible and uncertainties in the data are probably dominated by analytical artifacts as mentioned above, which cannot be accurately assessed based on our obtained data.

    The data comparisons shown in Fig. A7 suggest that the Al/Si and Mg/Si ratios in SiC grains determined based on NanoSIMS measurements of SRM 610 are generally overestimated, *i.e.*, RSF values are underestimated, by factors ranging from two to five. First of all, it is noteworthy that the X grain data from Liu et al. (2018b) in Fig. A7 were obtained with a duoplasmatron O⁻ source at a worse spatial resolution than the grain data from this study. As discussed in the main text, this probably means that the X grain data sampled an average of 45% Al contamination (Section 2), which roughly accounts for the larger slope (by a factor of ~2) determined by the X grain data in Fig. A7a. We excluded MS grain M1-A4-G476-1 and X grain M2-A2-G1036 (highlighted by black arrows) in obtaining the linear fits, because (1) the MS grain was quite small (0.47 μm) and had significant Al contamination that could not be completely excluded by selecting a small ROI (Fig. A5) and (2) the NanoSIMS data suggests that the X grain had an Al/Si ratio of 0.6, which is too high for a SiC. The high Al/Si ratio of the X grain was probably caused by Al contamination, since the $^{26}$Al/$^{27}$Al ratio inferred based on its NanoSIMS data (0.21±0.01) is a factor of 3–4 lower than those based on its SEM-EDX (0.93±0.11) and TEM-EDX data (0.77±0.08). The lower $^{26}$Al/$^{27}$Al ratio inferred by the NanoSIMS data suggests that ~70% of the sampled Al signals by NanoSIMS were from contamination instead of being intrinsic signals. In other words, the intrinsic Al/Si$_{NanoSIMS}$ ratio for the X grain should be a factor of around three lower, and the corrected data point falls within the 95% confidence band for the linear fit in Fig. A7a. By excluding these two grains, its respective Pearson's correlation coefficients (*R*) in Fig. A7a increase from 0.85 to 0.91 for the MS grains and 0.88 to 0.95 for the X grains. Second, the NanoSIMS X grain data in Fig. A7b are less likely to have suffered from significant Mg contamination. This is because for instance, Hoppe et al. (2015) reported that the Mg/Si ratios of all their studied MS grains lie below 0.003 (*i.e.*, the maximum level of Mg contamination) based on NanoSIMS analyses with a duoplasmatron source. In comparison, all but one X grain from Liu et al. (2018b) show Mg/Si$_{NanoSIMS}$ above 0.003, thereby suggesting that the Mg signals are dominated by $^{26}$Mg and that the effect of Mg contamination should be negligible in skewing the correlation shown in Fig. A7b. The X grain data in Fig. A7b show a strong linear correlation with a correlation coefficient *R* value of 0.95 (gray dashed line). However, the linear fit is dominantly controlled by the two grains with

the highest Mg/Si ratios (highlighted by black arrows), and many grains with low Mg/Si ratios lie below the 95% confidence band for the linear fit. As a test, we excluded these two grains and obtained a new linear fit (blue dotted line, $R = 0.90$), which includes more grains in the lower range within the confidence band. Given that the two slopes agree with each other within uncertainties, to be conservative, we adopted the shallower slope (2.43) for correcting the NanoSIMS Mg/Si data for presolar SiC (Table 1, Fig. 2b). In summary, according to the MS grain data in Fig. A7a and the X grain data in Fig. A7b, the RSF values of Al/Si and Mg/Si should be revised to 5.6±1.6 (95% confidence) and 7.5±2.4, respectively. The revised Mg/Si RSF value, 7.5, was adopted for calculating the Mg/Si ratios reported in Table 1 and Fig. 2b. For calculating the $^{26}Al/^{27}Al$ ratio, we still adopted the RSF value of 1.2 for Mg/Al based on the NanoSIMS analyses of Burma spinel, because this value agrees with the revised RSF value (1.3), which, however, has a much larger error due to the uncertainties in the linear fits in Fig. A7.

Finally, note that the SEM-EDX Mg/Si and Al/Si data of grain M2-A1-G674 are both 40% lower than its corresponding TEM-EDX data. The differences for grain M2-A1-G674 appear to be caused by some SEM-EDX analytical artifact, *e.g.*, topography, as the TEM-EDX Mg/Si data agree better with the linear fit shown in Fig. A7b. However, it is noteworthy that the TEM-EDX data for grain M2-A1-G674 show large Mg, Al heterogeneities across the TEM thin section (Singerling et al. 2021). The observed heterogeneities could be responsible for the inconsistencies between the EDX data and the linear fit in Fig. A7b and also between the TEM-EDX and SEM-EDX data, because the SEM-EDX data represent the grain's bulk composition while the NanoSIMS and TEM-EDX analyses only sampled a thin slice of the grain.

The Astrophysical Journal LettersLiu, N, Stephan, T., Boehnke, P., et al. 2017b, ApJL, 844, L12

Liu, N., Stephan, T., Boehnke, P., et al. 2018a, ApJ, 855, 144

Liu, N., Nittler, R. L., Alexander, C. M. O'D., and Wang J. 2018b, SciAdv, 4, eaao1054

Liu, N., Steele, A., Nittler, L. R., et al. 2017c, MAPS, 52, 2550

Liu, N., Stephan, T., Cristallo, S., et al. 2019, ApJ, 881, 28

Lodders, K., and Fegley, B. Jr. 1995, Meteoritics, 30, 661

Malherbe, J., Penen, F., Isaure, M.-P., et al. 2016, Anal. Chem., 88, 7130

Marty, B., Chaussidon, M., Wiens, R. C., Jurewicz, A. J. G., and Burnett, D. S. 2011, Science, 332, 1533

Morgan, D. H, Cannon, R. D., Hatzidimitriou, D., and Croke, B. F. W. 2003, MNRAS, 341, 534

Nittler, L. R., and Alexander, C. M. O' 2003, GeCoA, 67, 4961

Palmerini, S., La Cognata, M., Cristallo, S., and Busso, M. 2011, ApJ, 729, 3

Palmerini, S., Trippella, O., and Busso, M. 2017, MNRAS, 467, 1193

Schmidt, D. R., Woolf, N. J., Zega, T. J., and Ziurys, L. M. 2018, Nature, 564, 378

Singerling, S. A., Liu, N., Nittler, L. R., et al. 2021, ApJ, 913, 90

Straniero, O., Gallino, R., and Cristallo, S. 2006, NuPhA., 777, 311

Straniero, O., Imbriani, G., Strieder, F., et al. 2013, ApJ, 763, 100

Stephan, T., Trappitsch, R., Hoppe, P., et al. 2019, ApJ, 877, 101

Stephan, T., Bose, A., Boujibar, A., et al. 2021, LPI, 52, 2358

Ventura, P., Di Criscienzo, M., Carin, R., and D'Antona, F. 2013, MNRAS, 431, 3642

Wasserburg, G. J., Boothroyd, A. I., and Sackmann, I.-J. 1995, ApJ, 447, L37

Yang, X., Chen, P., and He, J. 2004, A&A, 414, 1049

Zinner, E. 2014, in Treatise on Geochemistry, Vol. 1, ed. A. M. Davis, (Elsevier, Oxford), 181

Zinner, E., Nittler, L. R., Gallino, R., et al. 2006, ApJ, 650, 350

Zinner, E., Amari, S., Guinness, R., et al. 2007. GeCoA, 71, 478620

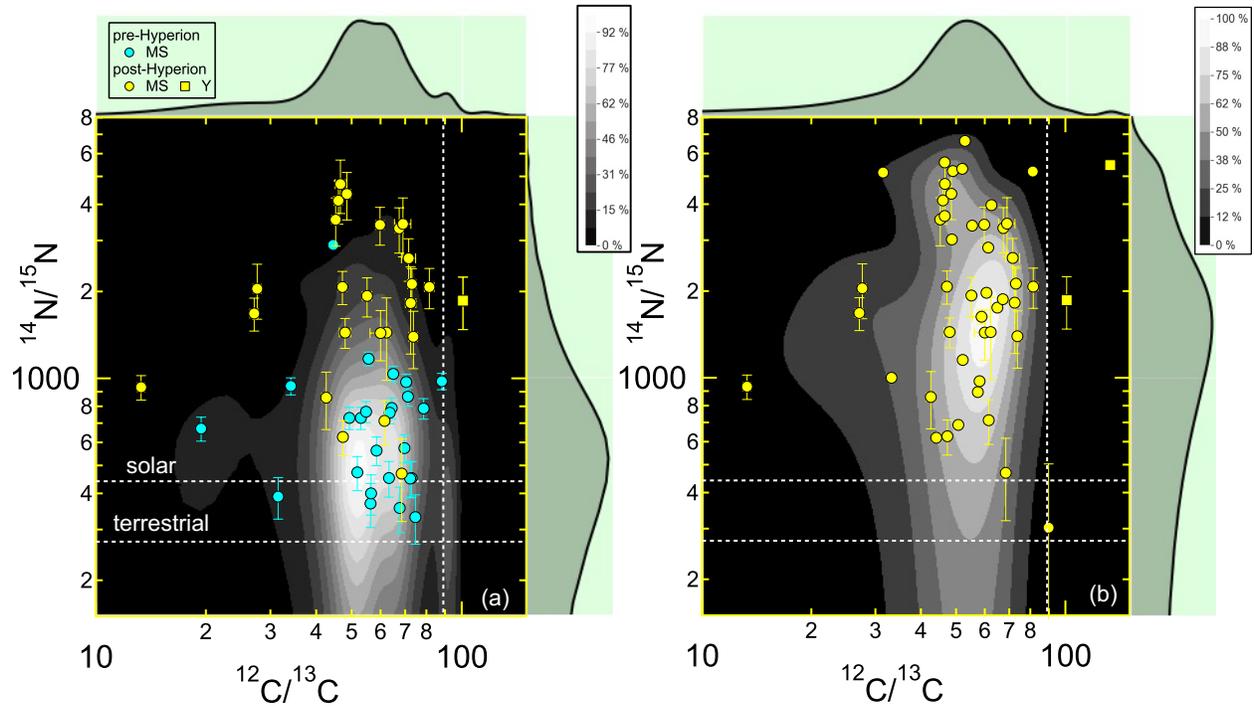

**Figure 1.** Plots of $^{14}N/^{15}N$ versus $^{12}C/^{13}C$. (a) The pre- and post-Hyperion data for 25 MS and Y grains from this study are compared to the probability distribution (in grayscale colormap) of MS, Y, and Z grain data (1σ errors < 2 for $^{12}C/^{13}C$ and <10% for $^{14}N/^{15}N$) from the literature (PGD_SiC_2021-01-10). (b) The post-Hyperion data for our 47 MS and Y grains (Table 1) are overlaid on their probability distribution (in grayscale colormap). Unless otherwise noted, dashed lines represent terrestrial values.

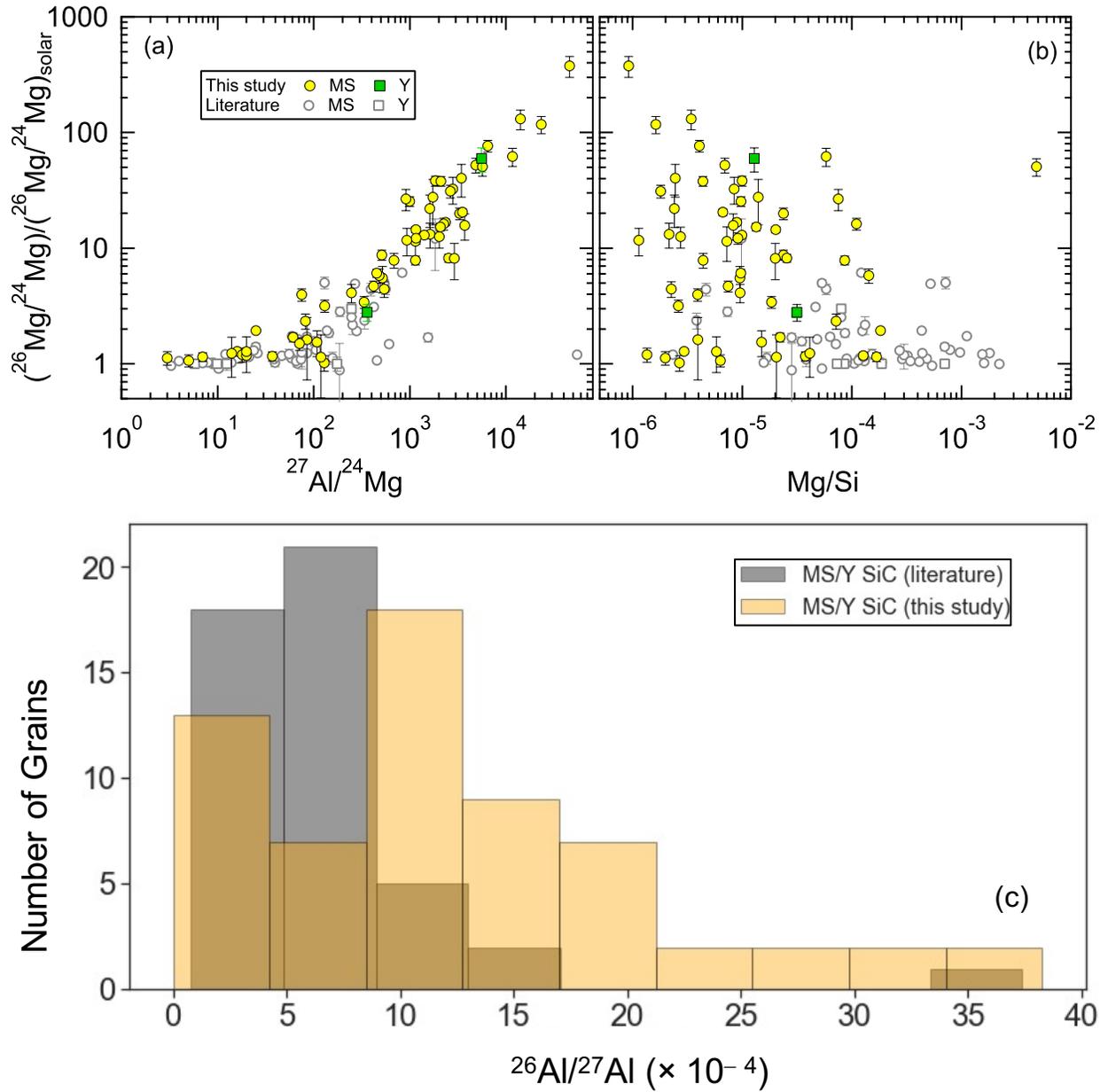

**Figure 2.** Plots comparing MS and Y SiC grains from this study and Huss et al. (1997) for their Mg-Al isotopic systematics and elemental ratios.

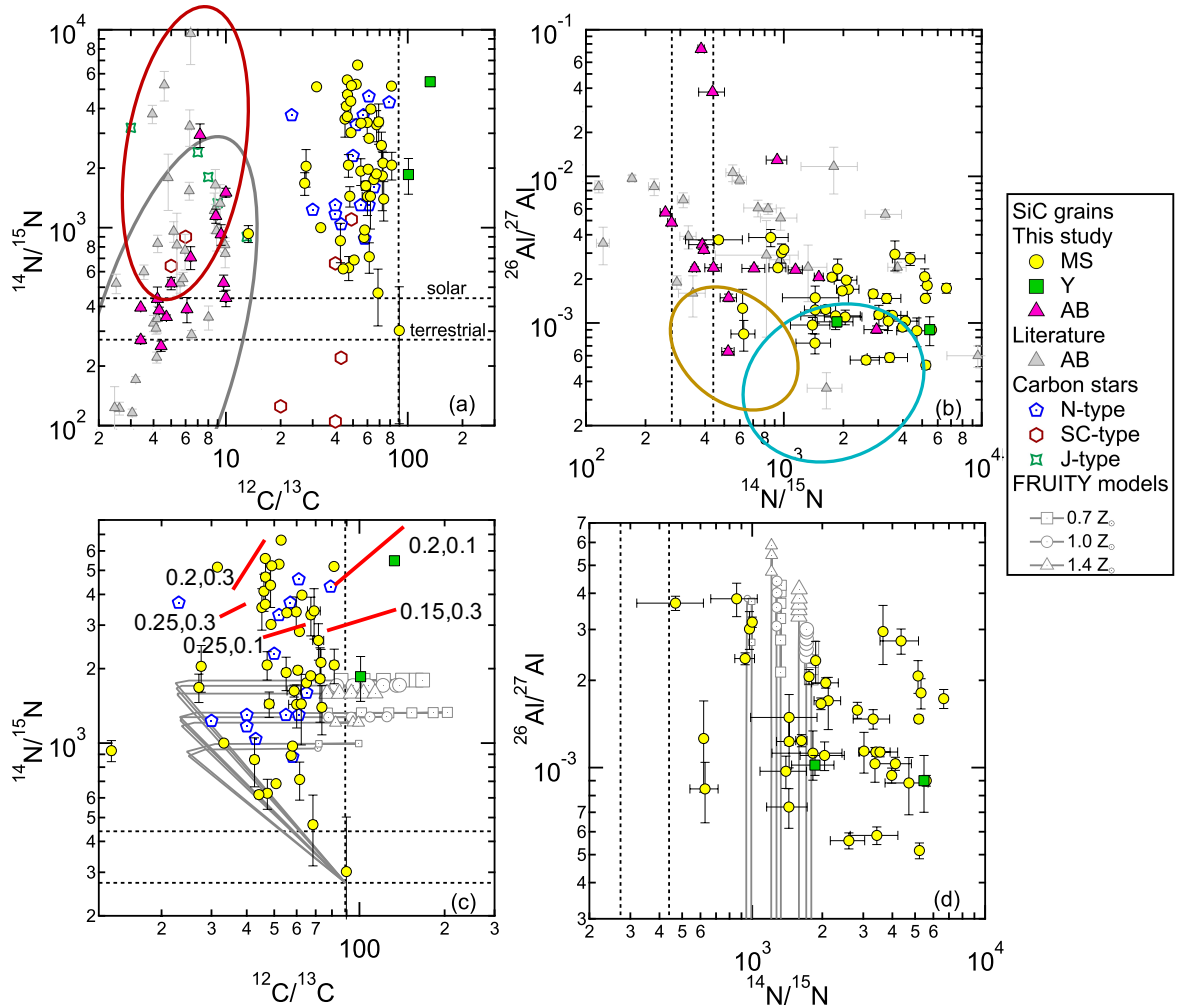

**Figure 3.** Plots comparing MS, Y, and AB SiC grains with observations for different types of carbon stars (Hedrosa et al. 2013; stars with upper limits reported for $^{14}N/^{15}N$ are not plotted) and FRUITY AGB models. The two groups of AB grains (fuchsia and grayscale ellipses) defined by B21 based on DB4 are given in (a) for comparison with AB grains from this study and Hoppe et al. (2019). The two groups of MS and Y grains (brown and blue ellipses) defined by B21 based on DB5 are shown in (b) for comparison with our MS/Y grains. In (c) and (d), FRUITY stellar models are shown as gray lines with symbols with the symbols representing C-rich phase during which SiC could condense; the small, medium, large symbol sizes correspond to 1.5 $M_\odot$, 2.0 $M_\odot$, and 3.0 $M_\odot$ stars, respectively. The CBP calculations of Palmerini et al. (2011) based on the FRUITY stellar model for a 2 $M_\odot$, $Z_\odot$ AGB star are shown in (c) as red lines with the two numbers denoting the values adopted for the two parameters in the CBP model (see the main text for details).

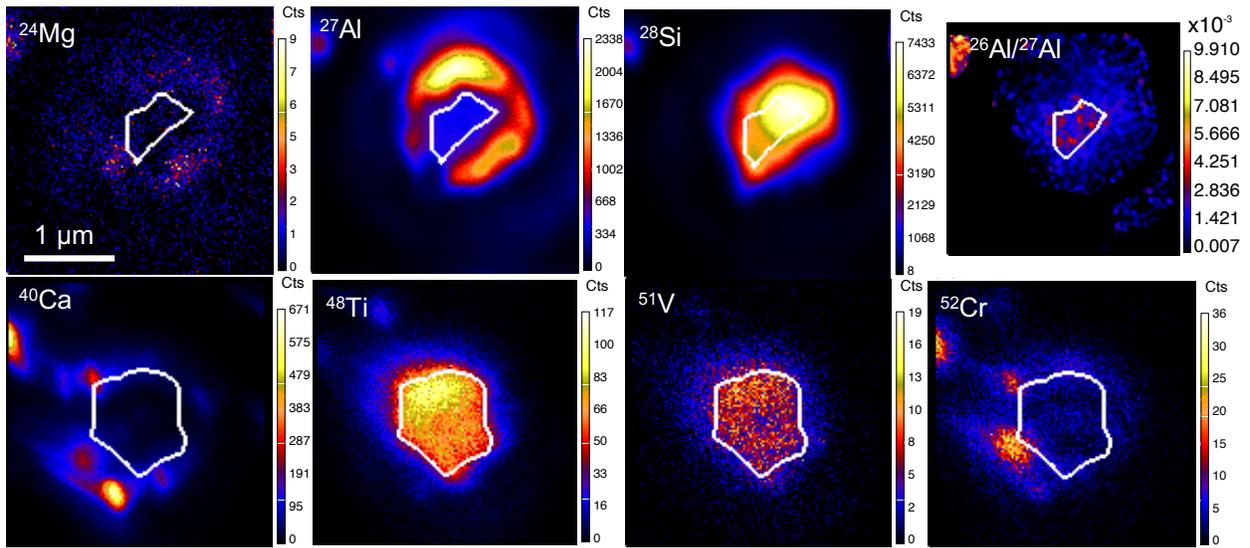

**Figure A1**. NanoSIMS ion (and inferred initial $^{26}$Al/$^{27}$Al) images of grain M1-A4-G508. The white lines highlight the contours of less contaminated grain regions (for Mg, Al, Ca, and Cr). Note that the ion images in the upper and lower panels were taken during two different analytical sessions.

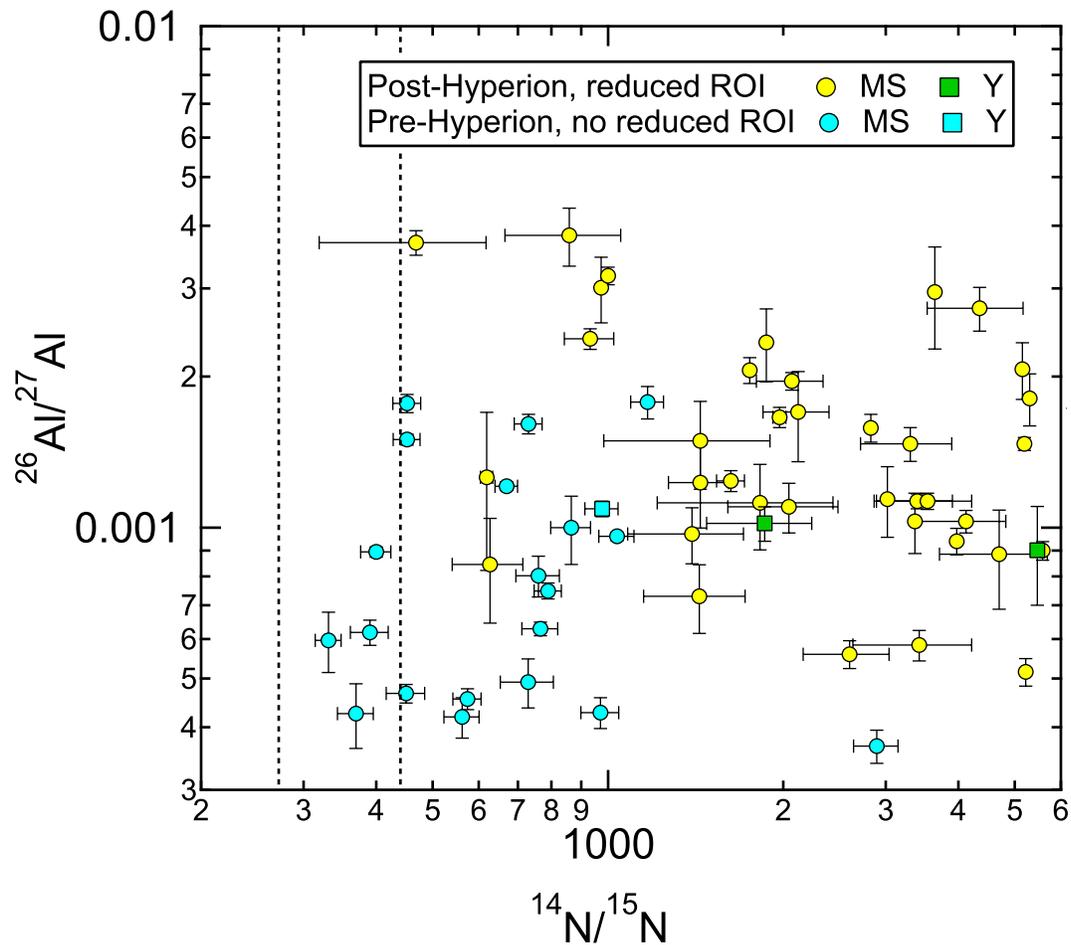

**Figure A2**. Plot of $^{26}$Al/$^{27}$Al versus $^{14}$N/$^{15}$N comparing the two sets of data for MS and Y grains from this study. Reduced ROI and no reduced ROI denote the cases of deducing initial $^{26}$Al/$^{27}$Al by excluding adjacent Al contamination using a small ROI and by including all Al signals contained within the SiC grain, respectively.

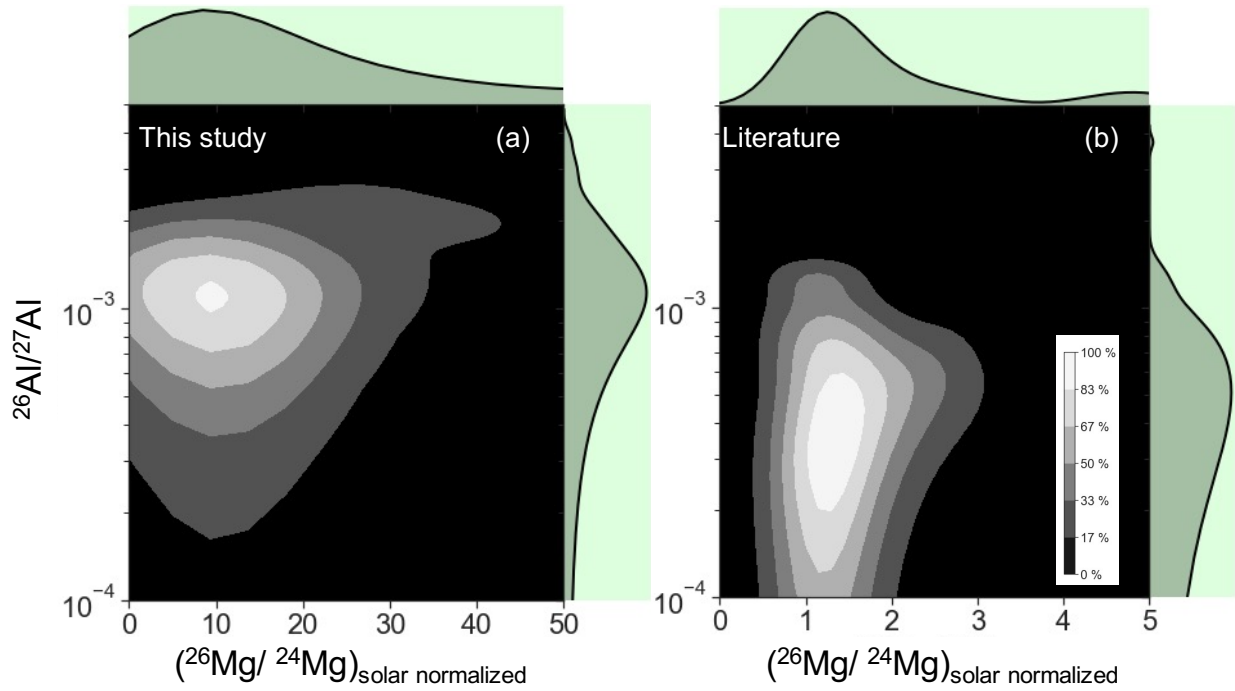

**Figure A3.** The probability distributions of $^{26}$Al/$^{27}$Al versus $^{26}$Mg/$^{24}$Mg normalized to the solar value for MS and Y SiC grain data from this study in (a) and Huss et al. (1997) in (b).

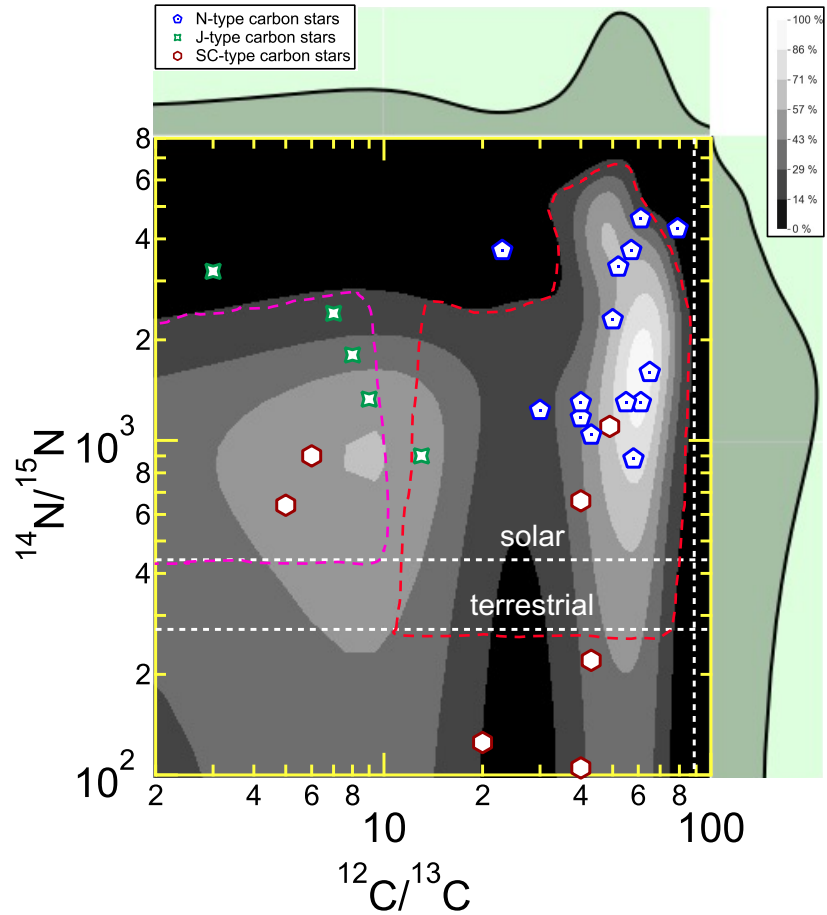

**Figure A4**. Observations for different types of carbon stars as shown in Fig. 3 are overlaid on the probability distribution (in grayscale colormap) of the post-Hyperion data for our 47 MS and Y grains (red dashed line region) and the pre-Hyperion data for our seven AB2 grains (fuchsia dashed line region).

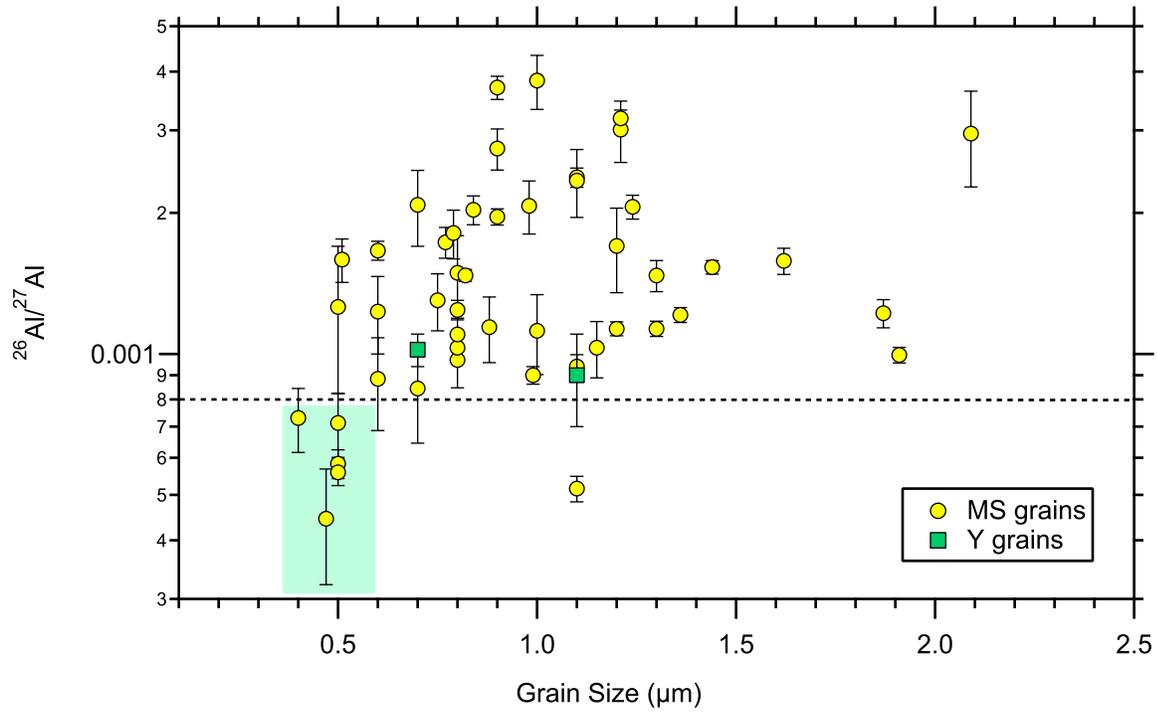

**Figure A5**. Plot of $^{26}$Al/$^{27}$Al ratio versus grain size for the MS and Y grains from this study.

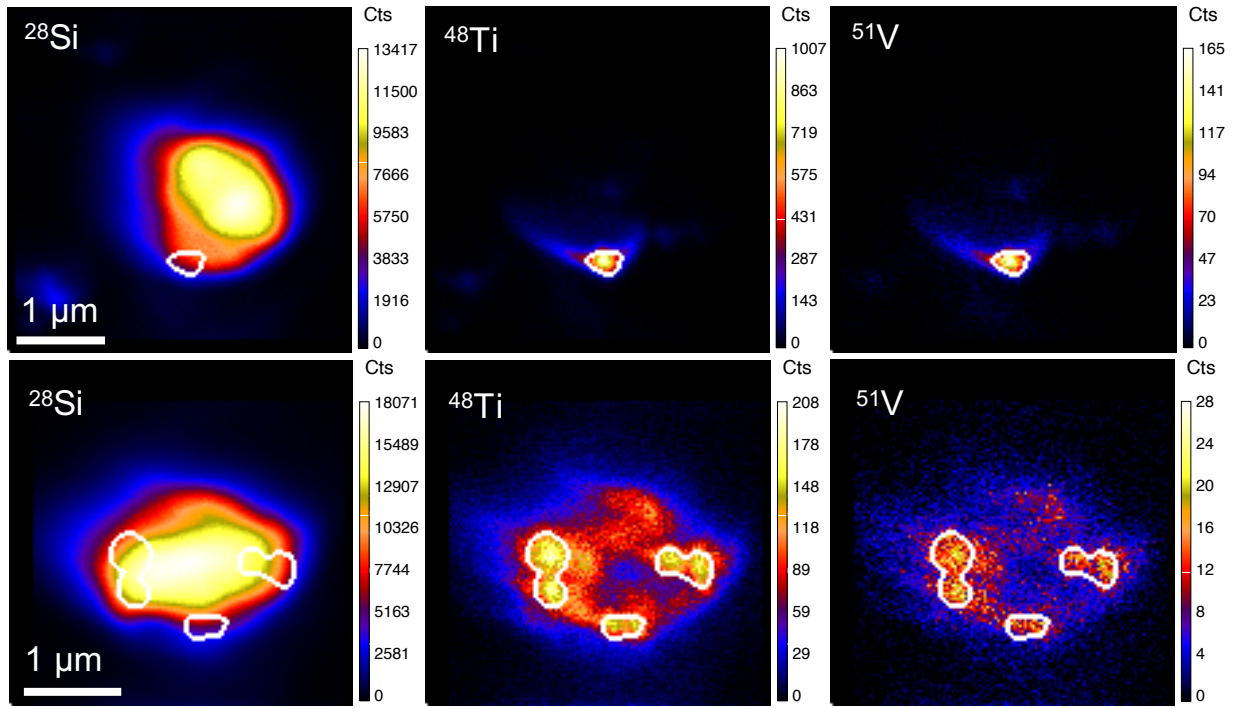

**Figure A6**. NanoSIMS ion images of MS grains M1-A2-277 (upper panel) and M1-A3-338 (lower panel). The white contours highlight Ti,V-hotspots within the grains.

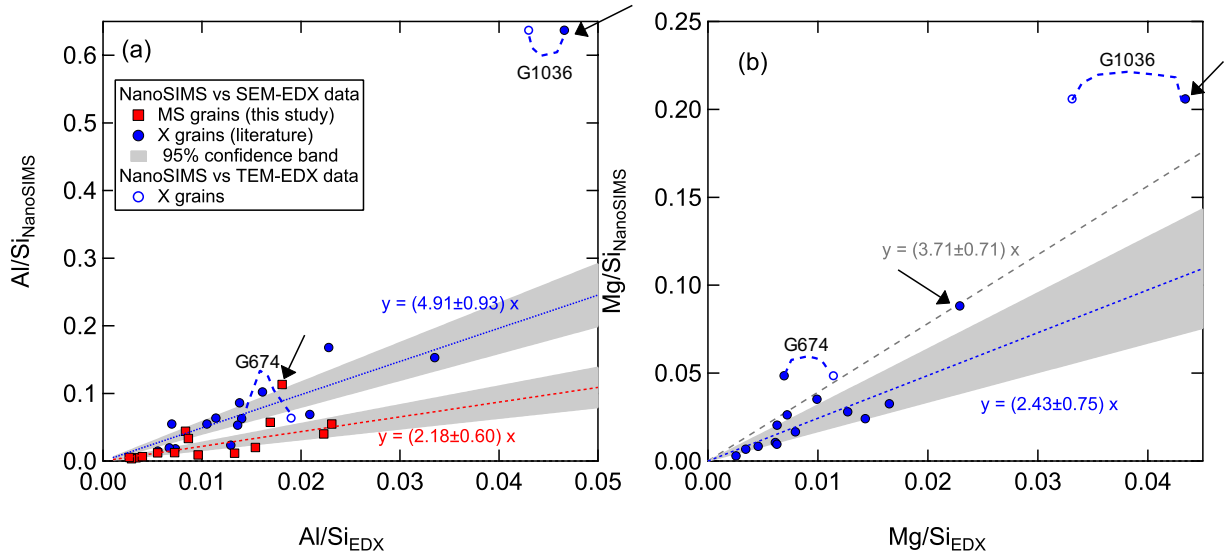

**Figure A7.** (a) Plot of Al/Si ratios determined by NanoSIMS versus those by SEM- and TEM-EDX for MS grains from this study and X grains from Liu et al. (2018b). A blue dashed curve connects TEM-EDX (open symbols) and SEM-EDX data (filled symbols) for the same grain. Black arrows denote grains that were excluded from the respective linear fits. (b) The same as (a) but for Mg/Si ratios. The grayscale dashed line represents the linear fit to all the grain data, and the blue dotted line is the linear fit to all but two grains (highlighted by black arrows).

**Table 1.** NanoSIMS Isotope Data of Presolar SiC Grains. All data are reported with 1σ errors.

| Grain | Type | Size (μm) | $^{12}C/^{13}C^a$ | $^{14}N/^{15}N$ | $\delta^{29}Si$ (‰) | $\delta^{30}Si$ (‰) | $\delta^{25}Mg$ (‰) | $\delta^{26}Mg$ (‰) | $^{26}Al/^{27}Al$ (×10$^{-3}$) | Mg/Si$^b$ (×10$^{-4}$) | Al/$^{24}$Mg |
|---|---|---|---|---|---|---|---|---|---|---|---|
| M1-A3-G203 | MS | 0.92 | 61.5±1.4 | 713±125 | 25±16 | 55±11 | 111±102 | 148±99 | | 1.68 | 7 |
| M1-A3-G220 | MS | 0.93 | 68.5±1.9 | 468±149 | 27±19 | 6±16 | −42±42 | 934±60 | 3.70±0.21 | 1.83 | 25 |
| M1-A3-G226 | MS | 0.69 | 47.3±1.1 | 627±87 | −13±17 | −14±15 | 143±546 | 4513±1412 | 0.84±0.20 | 0.10 | 519 |
| M1-A3-G243 | MS | 1.33 | 67.4±1.7 | 3309±590 | 37±15 | 64±9 | 270±250 | 7713±881 | 1.47±0.11 | 0.24 | 511 |
| M1-A3-G277 | MS | 1.22 | 59.8±1.5 | 3400±508 | 76±18 | 77±13 | 928±480 | 75462±8776 | 1.13±0.04 | 0.04 | 6475 |
| M1-A3-G338 | MS | 1.26 | 45.2±1.1 | 3539±674 | 47±18 | 49±14 | 39±152 | 13469±876 | 1.13±0.04 | 0.2 | 1156 |
| M1-A4-G476-1 | MS | 0.47 | 69.1±3.6 | 3425±790 | 106±18 | 67±14 | −118±179 | 6847±693 | 0.58±0.04 | 0.86 | 1141 |
| M1-A4-G476-2 | MS | 0.46 | 71.6±3.2 | 2602±438 | 124±21 | 67±23 | 306±352 | 18982±2352 | 0.59±0.04 | 0.24 | 3298 |
| M1-A4-G508 | MS | 1.06 | 13.3±0.3 | 932±91 | −32±15 | −10±10 | 449±251 | 24358±2471 | 2.38±0.11 | 0.10 | 995 |
| M1-A4-G612 | MS | 0.75 | 73.7±2.0 | 1394±314 | 50±17 | 40±13 | 18±289 | 4782±824 | 0.97±0.12 | 1.43 | 479 |
| M1-A5-G657 | MS | 0.76 | 62.3±1.5 | 1441±457 | 47±16 | 41±11 | 4714±3346 | 26566±11692 | 1.49±0.30 | 0.02 | 1735 |
| M1-A5-G662 | Y | 0.70 | 100.8±3.2 | 1858±382 | 16±11 | 28±8 | 321±350 | 58676±14120 | 1.02±0.08 | 0.13 | 5567 |
| M1-A5-G754 | AB1 | 0.71 | 3.4±0.1 | 271±12 | 3±16 | 3±12 | 228±95 | 22121±747 | 4.84±0.04 | 0.53 | 1954 |
| M1-A5-G950 | MS | 0.56 | 46.6±1.5 | 4702±987 | 100±17 | 95±12 | 67±803 | 10435±3764 | 0.89±0.20 | 0.07 | 1145 |
| M1-A5-G953 | MS | 0.74 | 81.5±2.5 | 2072±331 | −5±17 | 13±14 | 259±147 | 162±106 | | 0.22 | 37 |
| M1-A5-G960 | MS | 0.9 | 48.5±1.2 | 4353±815 | −35±15 | 16±10 | 379±668 | 25616±5567 | 2.74±0.27 | 0.03 | 909 |
| M1-A5-G1147 | MS | 0.41 | 60.0±3.8 | 1436±284 | 72±17 | 66±11 | 1154±917 | 12194±3213 | 0.73±0.11 | 0.14 | 1623 |
| M1-A5-G1205 | X | 0.51 | 106.8±1.4 | 55±2 | −185±27 | −323±25 | 246±201 | 121222±7209 | 94.89±1.47 | 2.58 | 573 |
| M1-A5-G1211 | MS | 0.78 | 46.0±1.2 | 4120±703 | 42±18 | 44±15 | 208±458 | 51320±7662 | 1.03±0.05 | 0.07 | 4858 |
| M1-A5-G1216 | MS | 0.96 | 72.5±2.4 | 1824±609 | 21±17 | 16±13 | −127±108 | 700±151 | 1.12±0.21 | 0.38 | 61 |
| M1-A5-G1381 | MS | 1.19 | 73.1±1.9 | 2122±276 | 17±16 | 17±11 | 24±57 | 280±62 | 1.70±0.35 | 0.75 | 16 |
| M1-A6-G096 | MS | 1.04 | 55.1±1.6 | 1930±298 | −29±18 | −12±16 | 1455±540 | 543±390 | | 0.08 | 108 |
| M1-A6-G099 | MS | 0.88 | 47.2±1.2 | 2071±272 | 90±18 | 91±14 | 311±316 | 37194±3768 | 1.96±0.08 | 0.10 | 1840 |
| M1-A6-G106-1 | MS | 1.05 | 42.6±1.2 | 858±193 | −15±19 | 8±17 | −217±193 | 2962±489 | 3.83±0.51 | 0.15 | 75 |
| M1-A6-G106-2 | MS | 0.67 | 27.1±0.7 | 1679±220 | 16±26 | 0±28 | 1000±845 | 234±465 | | 0.10 | 14 |
| M1-A6-G313 | AB1 | 0.36 | 10.0±0.1 | 441±42 | −57±55 | −20±71 | 19±128 | 5016±364 | 2.38±0.12 | 2.36 | 1142 |
| M1-A6-G428 | MS | 0.75 | 27.6±0.7 | 2045±439 | 114±17 | 105±12 | 1999±1753 | 31558±8474 | 1.10±0.13 | 0.04 | 2799 |
| M1-A6-G501 | AB2 | 0.54 | 7.2±0.1 | 2945±404 | 43±23 | 65±24 | 510±545 | 18814±3187 | 0.90±0.05 | 0.35 | 9779 |
| M1-A6-G506 | AB2 | 0.3 | 9.7±0.1 | 526±50 | 2±31 | −1±36 | −84±103 | 508±126 | 1.49±0.09 | 17.00 | 151 |
| M1-A6-G594 | AB1 | 0.48 | 4.4±0.1 | 252±15 | 89±32 | 68±35 | −28±75 | 8424±316 | 5.67±0.09 | 4.61 | 704 |
| M1-A6-G605 | MS | 0.65 | 48.0±1.7 | 1441±170 | 122±20 | 97±16 | 318±380 | 3120±739 | 1.23±0.23 | 0.41 | 247 |
| M1-A9-G509 | AB2 | 0.83 | 9.4±0.1 | 926±111 | −67±36 | −61±44 | 170±197 | 10650±859 | 12.95±0.18 | 1.17 | 404 |
| M1-A9-G1012 | X | 0.44 | 103.2±0.9 | 43±1 | −196±17 | −350±13 | 76±75 | 103865±2705 | 271.60±2.37 | 21.26 | 190 |
| M2-A2-G152 | AB2 | 0.26 | 6.4±0.1 | 708±94 | 23±49 | −13±59 | −191±845 | 263678±80933 | 2.36±0.09 | 0.02 | 48053 |
| M2-A2-G621 | AB1 | 0.42 | 6.1±0.1 | 387±56 | −13±32 | 39±39 | 482±925 | 51306±13166 | 3.43±0.25 | 0.04 | 7760 |
| M2-A2-G820 | X | 0.52 | 220.0±7.8 | | −274±11 | −393±13 | 96±62 | 39824±921 | 205.37±1.79 | 65.71 | 89 |
| M2-A2-G992 | X | 0.56 | 283.0±15.3 | | −339±15 | −566±16 | 172±178 | 308361±16534 | 134.70±1.07 | 6.02 | 699 |
| M2-A2-G1129 | AB2 | 0.38 | 8.8±0.2 | 1146±94 | 199±9 | 142±10 | −134±234 | 88402±7546 | 2.31±0.06 | 0.12 | 18614 |
| M2-A4-G1510 | X | 0.48 | 170.4±4.9 | | −298±10 | −432±12 | 76±158 | 1225817±60657 | 331.50±1.59 | 1.31 | 1617 |
| M3-G324 | X | 0.59 | 169.7±2.0 | 56±1 | −230±8 | −159±9 | −123±129 | 16357±926 | 39.03±0.19 | 4.00 | 182 |
| M3-G398 | AB1 | 0.43 | 4.3±0.1 | 382±14 | −38±8 | −7±8 | 89±85 | 24542±787 | 73.95±1.20 | 7.26 | 386 |
| M3-G472 | AB1 | 0.61 | 4.2±0.1 | 436±65 | 35±15 | 47±18 | −40±94 | 135628±4517 | 37.64±0.26 | 1.99 | 1707 |
| M3-G1154 | AB2 | 0.44 | 5.0±0.2 | 523±37 | 11±10 | 58±10 | −121±97 | 344±113 | 0.64±0.03 | 1.38 | 301 |
| M3-G1507 | AB2 | 0.47 | 10.0±0.1 | 1503±79 | −34±8 | −13±10 | 145±195 | 27229±1841 | 2.06±0.03 | 0.34 | 5428 |
| M3-G1607 | AB1 | 0.45 | 4.7±0.1 | 354±9 | 39±10 | 18±13 | −25±119 | 11600±618 | 2.37±0.05 | 2.48 | 2248 |
| M4-A1-1 | MS*$^c$ | 2.95 | | | 106±20 | 96±29 | −288±371 | 200±168 | | 0.01 | 18 |
| M4-A1-2 | MS* | 1.17 | | | 147±21 | 107±29 | 129±364 | 619±891 | | 0.04 | 85 |
| M4-A1-5 | MS* | 1.26 | | | −42±17 | 13±27 | −91±117 | 65±128 | | 0.06 | 5 |
| M4-A1-6 | MS* | 0.84 | | | 91±20 | 63±28 | −203±174 | 5081±484 | 2.03±0.14 | 0.10 | 452 |
| M4-A2-1 | MS* | 1.36 | | | 68±20 | 73±28 | −302±188 | 15723±1030 | 1.21±0.04 | 0.09 | 2344 |
| M4-A2-2 | MS | 1.24 | 65.0±0.3 | 1752±8 | 74±20 | 80±28 | −210±313 | 30069±3955 | 2.06±0.12 | 0.02 | 2637 |
| M4-A2-3 | MS | 1.15 | 55.4±0.4 | 3369±24 | 41±19 | 90±29 | −448±181 | 11551±2586 | 1.03±0.14 | 0.03 | 2021 |
| M4-A2-4 | X | 1.73 | | | −301±13 | −471±14 | −162±164 | 10846858±2058167 | 627.86±1.25 | 0.08 | 3106 |
| M4-A2-5 | MS* | 1.02 | | | 24±19 | 55±28 | 32±50 | 181±82 | | 1.27 | 20 |
| M4-A3-3 | Y | 1.10 | 132.8±1.5 | 5466±49 | 21±19 | 47±28 | −42±97 | 1792±466 | 0.90±0.20 | 0.27 | 359 |
| M4-A3-4 | MS* | 1.44 | | | 17±19 | 29±27 | 78±98 | 11984±646 | 1.53±0.05 | 0.10 | 1409 |
| M4-A4-1 | MS | 1.10 | 67.2±0.4 | 1872±25 | 54±19 | 50±28 | 271±433 | 20978±6825 | 2.34±0.39 | 0.02 | 1616 |
| M4-A4-2 | MS | 0.82 | 81.3±1.0 | 5191±60 | 43±19 | 37±27 | 45±641 | 375282±77589 | 1.47±0.04 | 0.01 | 45984 |
| M4-A4-3 | MS | 0.99 | 46.5±0.2 | 5584±21 | 31±19 | 37±28 | 933±814 | 116256±19899 | 0.90±0.04 | 0.02 | 23291 |
| M4-A4-4 | MS | 1.36 | 57.4±0.3 | 892±3 | 110±20 | 97±29 | 495±435 | 125±152 | | 0.02 | 3 |
| M4-A4-5 | MS | 1.10 | 49.0±0.2 | 5218±22 | 2±19 | 9±27 | 158±259 | 7218±653 | 0.52±0.03 | 0.26 | 2524 |
| M4-A4-6 | MS | 0.77 | 52.9±0.4 | 6632±50 | −32±18 | 25±27 | 119±235 | 11170±1343 | 1.73±0.13 | 0.09 | 1161 |
| M4-A4-7 | MS | 0.60 | 60.6±1.6 | 1973±51 | 29±19 | 22±27 | −134±456 | 130319±25533 | 1.66±0.08 | 0.03 | 14134 |
| M4-A4-8 | MS* | 0.50 | | | 6±19 | 37±28 | 579±462 | 14719±4000 | 0.71±0.11 | 0.08 | 3729 |
| M4-A4-9 | MS | 1.21 | 58.0±0.3 | 973±4 | 53±19 | 63±28 | 393±534 | 2169±384 | 3.01±0.45 | 0.03 | 130 |
| M4-A4-10 | MS* | 0.47 | | | 43±21 | 46±29 | −110±420 | 7173±2847 | 0.45±0.12 | 0.20 | 2906 |
| M4-A5-1 | MS | 1.62 | 61.3±0.3 | 2830±15 | 57±19 | 75±28 | 201±263 | 49690±8711 | 1.58±0.10 | 0.02 | 5676 |
| M4-A5-2 | MS | 0.98 | 31.5±0.1 | 5153±18 | 140±21 | 133±30 | −384±452 | 39385±12720 | 2.07±0.27 | 0.01 | 3434 |
| M4-A5-3 | MS | 0.50 | 44.1±1.1 | 619±15 | 143±21 | 136±30 | −71±80 | 495±189 | 1.26±0.44 | 1.10 | 71 |
| M4-A7-1 | MS* | 1.87 | | | 9±18 | 38±27 | −103±179 | 15202±1851 | 1.22±0.08 | 0.07 | 2253 |
| M4-A7-2 | MS* | 1.91 | | | −8±18 | 33±27 | −225±210 | 19468±1420 | 1.00±0.04 | 0.04 | 3529 |

| Grain | Type | | | | | | | | | |
|---|---|---|---|---|---|---|---|---|---|---|
| M4-A7-3 | MS | 1.21 | 33.2±0.2 | 1000±5 | 33±19 | 48±28 | −73±209 | 36981±3681 | 3.18±0.13 | 0.13 | 2097 |
| M4-A7-4 | MS | 0.80 | 58.8±3.3 | 1627±90 | 45±19 | 48±28 | 57±164 | 14261±1186 | 1.24±0.06 | 0.18 | 2078 |
| M4-A8-1 | MS* | 0.75 | | | 64±20 | 90±29 | −42±124 | 2422±401 | 1.30±0.18 | 0.58 | 335 |
| M4-A9-3 | MS | 1.10 | 62.6±0.3 | 3973±16 | 45±19 | 62±28 | 1670±796 | 60978±11107 | 0.94±0.06 | 0.03 | 11708 |
| M4-A9-4 | MS* | 0.50 | | | −8±19 | 19±28 | 144±177 | 20±157 | | 0.72 | 129 |
| M4-A10-2 | MS | 2.09 | 46.5±0.2 | 3645±15 | 155±21 | 122±30 | −58±176 | 1345±355 | 2.95±0.68 | 0.06 | 82 |
| M4-A10-3 | MS | 0.69 | 50.7±0.5 | 686±7 | −75±17 | 81±29 | 1±531 | 279±435 | | 0.02 | 20 |
| M4-A10-4 | MS | 0.88 | 48.7±0.3 | 3021±16 | −6±18 | 31±27 | −44±221 | 3425±689 | 1.14±0.18 | 0.04 | 540 |
| M4-A10-5 | MS | 0.79 | 52.0±0.5 | 5304±55 | 13±19 | 50±28 | 152±263 | 6863±1166 | 1.81±0.26 | 0.07 | 684 |
| M4-A10-6-1 | AB1 | 0.62 | 3.4±0.1 | 394±1 | −35±18 | 24±28 | 1883±1121 | 115237±39465 | 3.16±0.16 | 0.05 | 6565 |
| M4-A10-6-2 | X | 0.32 | | | −342±13 | −578±12 | 92±213 | 2560159±478185 | 267.50±1.10 | 0.33 | 1721 |
| M4-A10-7 | MS* | 0.51 | | | 57±20 | 48±29 | 144±339 | 3688±499 | 1.59±0.17 | 0.21 | 419 |
| M4-A10-8 | MS | 1.12 | 89.9±0.9 | 302±200 | 24±19 | 8±27 | 496±396 | 145±637 | | 0.01 | 119 |
| M4-A10-9 | MS* | 0.70 | | | 104±20 | 97±29 | 86±409 | 10706±3142 | 2.08±0.38 | 0.03 | 930 |

a: Carbon and nitrogen isotopic ratios for MS/Y grains reported here are those measured after the analyses of Mg-Al and Ti isotopes (i.e., post-Hyperion data).
b: The revised RSF value for Mg/Si based on SEM-EDX data, 7.5, was adopted for calculation (see Appendix B for details).
c: MS* denotes that the grain is classified as type MS solely based on its Si isotopic composition.

**Table 2.** Abundances and Chemical Characteristics of Carbon Stars

| Carbon Stars | Abundance | $^{12}C/^{13}C$ | $^{14}N/^{15}N$ | s-process enrichment |
|---|---|---|---|---|
| N-type | large majority (e.g., Lloyd Evans 2010) | 20–90 | 1000–5000 | yes |
| J-type | 10-15% in the Galaxy and the Large Magellanic Cloud (Abia & Isern 2000; Morgan et al. 2003) | ≤13 | 1000–3000 | no |
| SC-type | rare | 5–50 | 100–1000 | yes |

**Table A1.** NanoSIMS Isotope Data of Presolar SiC Grains. All the data are reported with 1σ errors.

| Grain | Type | $^{12}C/^{13}C$[a] | $^{14}N/^{15}N$ | $\delta^{46}Ti$ (‰) | $\delta^{47}Ti$ (‰) | $\delta^{49}Ti$ (‰) | $\delta^{50}Ti$ (‰) |
|---|---|---|---|---|---|---|---|
| M1-A3-G203 | MS | 67.7±0.3 | 356±16 | −34±23 | 7±23 | −2±12 | 27±8 |
| M1-A3-G220 | MS | 73.0±0.5 | 452±25 | 40±35 | −11±35 | | |
| M1-A3-G226 | MS | 56.3±0.4 | 369±26 | −43±13 | −18±11 | 39±12 | 62±8 |
| M1-A3-G243 | MS | 64.4±0.2 | 789±42 | −10±12 | −31±9 | 41±12 | 268±10 |
| M1-A3-G277 | MS | 65.0±0.4 | 1037±72 | 39±15 | 19±12 | 113±13 | 190±9 |
| M1-A3-G338 | MS | 56.5±0.4 | 400±24 | 2±12 | −21±8 | 51±12 | 192±9 |
| M1-A4-G476-1 | MS | 69.7±0.4 | 574±32 | 20±20 | −10±18 | 36±13 | 290±11 |
| M1-A4-G476-2 | MS | 72.3±0.7 | 450±34 | 4±17 | 9±16 | 137±14 | 86±9 |
| M1-A4-G508 | MS | 19.4±0.1 | 670±30 | −115±12 | −67±10 | 63±13 | −143±7 |
| M1-A4-G612 | MS | 70.5±0.5 | 971±72 | −31±16 | −24±14 | 141±17 | 224±18 |
| M1-A5-G657 | MS | 63.6±0.4 | 760±65 | −6±14 | −3±12 | 135±13 | 180±9 |
| M1-A5-G662 | Y | 88.2±0.6 | 977±64 | 1±15 | −16±13 | 98±13 | 197±9 |
| M1-A5-G950 | MS | 44.5±0.3 | 2895±255 | 64±13 | 24±8 | 139±13 | 218±9 |
| M1-A5-G953 | MS | 78.7±0.6 | 787±73 | −44±17 | −9±16 | 83±13 | 165±9 |
| M1-A5-G960 | MS | 53.0±0.2 | 730±40 | −97±11 | −65±8 | -25±11 | 76±8 |
| M1-A5-G1147 | MS | 58.5±0.3 | 562±39 | 65±23 | 27±22 | 36±12 | 116±9 |
| M1-A5-G1211 | MS | 54.8±0.4 | 766±54 | 48±16 | 9±13 | 31±13 | 93±10 |
| M1-A5-G1216 | MS | 74.7±0.4 | 331±17 | 27±15 | 1±12 | 79±13 | 236±9 |
| M1-A5-G1381 | MS | 71.2±0.4 | 865±68 | −13±15 | −19±12 | 111±13 | 155±9 |
| M1-A6-G096 | MS | 51.8±0.4 | 473±36 | | | 16±12 | 171±9 |
| M1-A6-G099 | MS | 63.3±0.3 | 452±24 | 69±16 | 18±13 | 36±12 | 196±9 |
| M1-A6-G106-1 | MS | 55.6±0.4 | 1170±76 | −9±34 | −33±35 | 4±62 | 127±67 |
| M1-A6-G106-2 | MS | 34.1±0.2 | 942±96 | | | 87±16 | 239±13 |
| M1-A6-G428 | MS | 31.5±0.2 | 390±29 | | | | |
| M1-A6-G605 | MS | 49.3±0.4 | 729±76 | 119±23 | 81±21 | 139±13 | 202±9 |
| M1-A5-G754 | AB1 | | | −90±14 | −62±12 | −24±11 | −86±7 |
| M1-A6-G313 | AB1 | | | −88±17 | −68±16 | −4±14 | 40±11 |
| M1-A6-G594 | AB1 | | | 71±18 | 18±16 | 175±14 | 110±8 |
| M3-G472 | AB1 | | | −4±51 | −67±52 | 70±13 | −3±8 |
| M3-G1607 | AB1 | | | | | 32±12 | −73±7 |
| M1-A6-G501 | AB2 | | | 91±18 | 14±15 | 178±14 | 288±10 |
| M1-A6-G506 | AB2 | | | | | 50±14 | 75±10 |
| M2-A2-G1129 | AB2 | | | 74±19 | 34±17 | −13±12 | |
| M3-G1154 | AB2 | | | −88±19 | −99±30 | 134±14 | 112±10 |
| M3-G1507 | AB2 | | | 109±16 | 51±12 | 155±14 | 232±9 |
| M2-A2-G820 | X | | | −36±50 | −48±52 | 269±15 | 210±10 |

a: Carbon and nitrogen isotopic ratios for MS/Y grains reported here are those measured before the analyses of Mg-Al and Ti isotopes (i.e., pre-Hyperion data).